\documentclass[12pt,a4paper]{article}
\usepackage[T2A]{fontenc}
\usepackage{amsfonts}
\usepackage{amsmath}
\usepackage{amssymb}
\usepackage[english]{babel}
\usepackage{graphicx}
\newtheorem{law1}{Lemma}
\newtheorem{law2}{Theorem}

\begin{document}
\title{Discrete mechanics: a kinematics for a particular case of causal sets}
\author{Alexey L. Krugly\thanks{Quantum Information Laboratory, Institute of Physics and Physical Technologies, Moscow, Russia; akrugly@mail.ru.}}
\date{} \maketitle
\begin{abstract}
The model is a particular case of causal set. This is a discrete model of spacetime in a microscopic level. In paper the most general properties of the model are investigated without any reference to a dynamics. The dynamics of the model is introduced in [arXiv: 1004.5077]. These two papers introduce a consistent description of the model.
\end{abstract}
\newpage
\tableofcontents
\newpage
\section{INTRODUCTION\label{I}}
The idea of building simple models to test the hypothesis of spacetime discreteness, or some aspect of it, is intriguing. One of such approach to quantum gravity is a causal set program (see e.g. \cite{Sorkin2005,Dowker2006,Henson2009,Wallden2010}. The central hypothesis is that on a microscopic scale, spacetime is a partially ordered set of purely discrete points. This idea is proposed by J. Myrheim \cite{Myrheim} and G. 't Hooft \cite{'t Hooft}.

A causal set is a pair ($\mathcal{C}$, $\prec$), where $\mathcal{C}$ is a set and $\prec$ is a binary relation on $\mathcal{C}$ satisfying the following properties ($x,\ y,\ z$ are general points in $\mathcal{C}$):
\begin{equation}
\label{eq:I1.1} x\prec x\qquad\textrm{(irreflexivity),}
\end{equation}
\begin{equation}
\label{eq:I1.2} \{x\mid(x\prec y)\wedge(y\prec x)\}=\emptyset \qquad \textrm{(acyclicity),}
\end{equation}
\begin{equation}
\label{eq:I1.3} (x\prec y)\wedge(y\prec z)\Rightarrow(x\prec z)\qquad \textrm{(transitivity),}
\end{equation}
\begin{equation}
\label{eq:I1.4} \mid\mathcal{A}(x, y)\mid<\infty\qquad\textrm{(local finiteness),}
\end{equation}
where $\mathcal{A}(x,\ y)$ is an Alexandrov set of the elements $x$ è $y$. $\mathcal{A}(x,\ y)=\{z\mid x\prec z \prec y\}$. The local finiteness means that the Alexandrov set of any elements is finite. Sets of elements are denoted by calligraphic capital Latin letters.

One of the goals of the causal set programme is to investigate the emergence of continuous spacetime as an approximation of some kind of causal sets. The main idea is a faithful embedding of causal sets \cite{BMSorkin}. Such causal set must possesses specific properties. But spacetime can emerge from causal set only after some coarse graining. The primordial causal set can be unfaithfully embedded \cite{1006.2320}. We can use this primordial causal set for the description of particles without any reference to spacetime. The simple particular model of the primordial causal set and its dynamics is introduced in \cite{1004.5077}. In this paper I investigate a kinematics of this model. It seems natural to follow the scheme that a theory is comprised of three components: kinematics, dynamics, and phenomenology. For causal sets, kinematics refers first of all to the kind of structure one has and its properties. The dynamics means what one might describe as the `equations of motion' of the causal set. The word `phenomenology' needs no definition.

In the next section I introduce a model. In section \ref{P} the most general properties are investigated. In section \ref{SG} I present the result that is important for a dynamics. In section \ref{A} the properties of slices are introduced. The slice is the discrete analog of a spacelike hypersurface. In section \ref{W} I prove the theorem that all slices have the same cardinality in this model. In section \ref{IO} I introduce some ideas that concern the definition of physical objects. In section \ref{LC} one property of light cones is introduced. In section \ref{CON} some open questions are discussed.

\section{A MODEL\label{M}}

There are a set of primordial indivisible objects. They have not an internal structure. Consequently, they itself have not any internal properties except one. They exist. The property ``existence'' can adopt two values: ``the primordial object exists'', and ``the primordial object does not exist''. There are two elementary processes. The first is a creation of a primordial object. This process changes the value of the property ``existence'' from ``the primordial object does not exist'' to ``the primordial object exists''. It is denoted by $\alpha_i$. The second is an annihilation of a primordial object. This process changes the value of the property ``existence'' from ``the primordial object exists'' to ``the primordial object does not exist''. It is denoted by $\beta_j$.

\begin{figure}
	\centering	
		\includegraphics[trim=8cm 18cm 8cm 9cm]{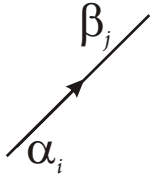}
	\caption{A chronon.}
	\label{fig:fig1}
\end{figure}

\begin{figure}
	\centering	
		\includegraphics[trim=8cm 18cm 8cm 7cm]{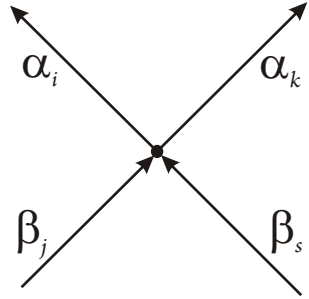}
	\caption{A x-structure.}
	\label{fig:fig2}
\end{figure}

In this model, any physical process is a finite network of finite elementary processes \cite{STC, STC2, STC3, STC4, FinkMcC1, Fink88}. Below a terminology of David Finkelstein is used. These two elementary processes are called monads \cite{STC4}. A propagation of the primordial object is simply an ordered pair of creation and annihilation. This process of propagation is called a chronon \cite{STC} (Fig.\ \ref{fig:fig1}).
The primordial object can be destroyed only by the interaction with another primordial object. The interaction of this second primordial object means the change of its state. Only one kind of change is possible. This is the annihilation of the second primordial object. Suppose the number of the primordial objects does not change. This is a fundamental conservation law. We have a simplest interaction process: two primordial objects are destroyed and two primordial objects are created. This process is called a tetrad \cite{FinkMcC1} or an x-structure \cite{Krugly2002} (Fig.\ \ref{fig:fig2}).
The x-structure consists of two monads of destruction and two monads of creation. Suppose any process can be divided into x-structures. This symmetric dyadic kinematics is called X kinematics \cite{Fink88}.

\begin{figure}
	\centering	
		\includegraphics[trim=8cm 17cm 8cm 7cm]{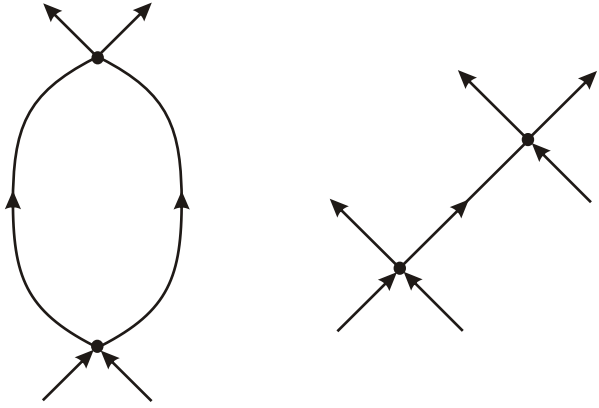}
	\caption{Connected d-graphs that consist of two x-structures.}
	\label{fig:fig3}
\end{figure}

\begin{figure}
	\centering	
		\includegraphics[trim=8cm 5cm 8cm 5cm]{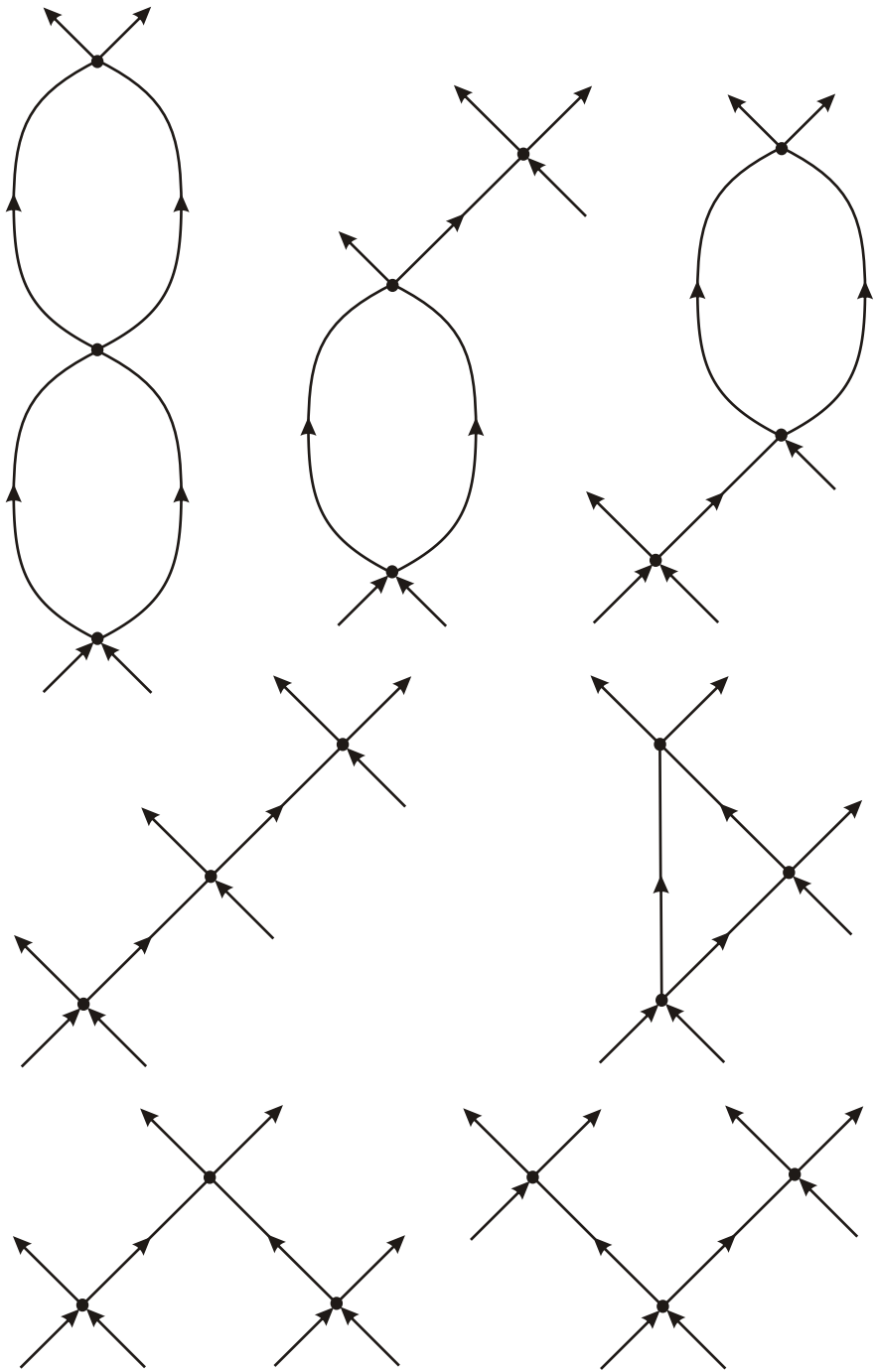}
	\caption{Connected d-graphs that consist of three x-structures.}
	\label{fig:fig4}
\end{figure}

Suppose there is a universal causal order of monads. Consider only finite sets of x-structures. Such set forms a structure. This structure is called d-graph \cite{1004.5077}. Consider simple examples. There are two connected d-graphs that consist of two x-structures (Fig.\ \ref{fig:fig3}) and seven connected d-graphs that consist of three x-structures (Fig.\ \ref{fig:fig4}). (The rigorous definition of a connected d-graph will be given in section \ref{SG}).
A d-graph is not a graph. By definition, a graph is a set of vertexes and a binary relation (edges) over this set. We cannot describe external lines as in Feynman diagrams. A d-graph is like a Feynman diagram. There are internal lines or edges. These are chronons. They consist of two monads. Monads are halves of edges \cite{FinkMcC1}. There are external monads. These external monads is useful for the description of dynamics \cite{1004.5077}.

Introduce the axiomatic approach to this model. Consider the set $\mathcal{G}$ of monads and a binary relation (an immediate causal priority) over this set. By $(\alpha_i\beta_j)$ denote an immediate causal priority relation of $\alpha_i$ and $\beta_j$. $\mathcal{G}$ satisfies the following axioms.
\begin{equation}
\label{eq:M1.1}
\forall\alpha_i(\exists !\beta_j(\alpha_i\beta_j))\vee(\not\exists\beta_j(\alpha_i\beta_j))\textrm{,}
\end{equation}
\begin{equation}
\label{eq:M1.2}
\forall\alpha_i(\not\exists \alpha_j(\alpha_i\alpha_j))\textrm{,}
\end{equation}
\begin{equation}
\label{eq:M1.3}
\forall\beta_j(\exists !\alpha_i(\alpha_i\beta_j))\vee(\not\exists\alpha_i(\alpha_i\beta_j))\textrm{,}
\end{equation}
\begin{equation}
\label{eq:M1.4}
\forall\beta_j(\not\exists \beta_i(\beta_i\beta_j))\textrm{.}
\end{equation}
These axioms describe a chronon. There is no more than one monad $\beta_j$ and does not exist the monad $\alpha _j$ which immediately causally follows any $\alpha_i$. There is no more than one monad $\alpha_i$ and does not exist the monad $\beta_i$ which immediately causally precedes any $\beta_j$. The pair $(\alpha_i\beta_j)$ is called a chronon or an edge. The monad is called internal iff it is included in a chronon. Otherwise the monad is called external.

The following axioms describe an x-structure.
\begin{equation}
\label{eq:M1.5}
\forall\alpha_i\exists !\alpha_j(\forall\beta_k(\beta_k\alpha_i)\Rightarrow(\beta_k\alpha_j))\textrm{,}
\end{equation}
\begin{equation}
\label{eq:M1.6}
\forall\beta_i\exists !\beta_j(\forall\alpha_k(\beta_i\alpha_k)\Rightarrow(\beta_j\alpha_k))\textrm{.}
\end{equation}
There is two and only two monads $\beta_i$ and $\beta_j$ which immediately causally precede any $\alpha_k$. There is two and only two monads $\alpha_i$ and $\alpha_j$ which immediately causally follow any $\beta_k$. Each monad is included in a unique x-structure.

By definition, two monads $\alpha_i$ and $\alpha_j$ are causally connected ($\alpha_i\prec\alpha_j$) iff there is the sequence $(\alpha_i\beta_k)(\beta_k\alpha_l)\dots(\beta_r\alpha_j)$. Such sequence of monads is called a saturated chain or a path. A causality is described by the following axiom.
\begin{equation}
\label{eq:M1.7}
\{\alpha_i|(\alpha_i\beta_k)(\beta_k\alpha_l)\dots(\beta_j\alpha_i)\}= \emptyset\textrm{.}
\end{equation}
We consider only finite sets of monads.
\begin{equation}
\label{eq:M1.8} |\mathcal{G}|<\infty\textrm{.}
\end{equation}
The set $\mathcal{G}$ of monads is d-graph if it satisfies these axioms.

\section{PROPERTIES OF D-GRAPHS\label{P}}

Consider a main properties of d-graphs.

\begin{law1}\label{L1}
\begin{equation}
\label{eq:P1.1}
\{\beta_i|(\beta_i\alpha_k)(\alpha_k\beta_l)\dots(\alpha_j\beta_i)\}= \emptyset\textrm{.}
\end{equation}
\end{law1}

Proof: Otherwise axiom (\ref{eq:M1.7}) is not satisfies for $\alpha_k$. $\Box$

The monad of any type is denoted by $\gamma_i$. The monad $\gamma_i$ may be $\alpha_i$ or $\beta_i$. By definition, two monads $\gamma_i$ and $\gamma_j$ are causally connected ($\gamma_i\prec\gamma_j$) iff there is the path $(\gamma_i\gamma_k)(\gamma_k\gamma_l)\dots(\gamma_r\gamma_j)$. By definition, put $\mathcal{A}(\gamma_i,\ \gamma_j)=\{\gamma_s| \gamma_i\prec \gamma_s\prec \gamma_i\}$. The set $\mathcal{A}(\gamma_i,\ \gamma_j)$ is called an Alexandrov set of $\gamma_i $ and $\gamma_j$. By definition, put $\mathcal{\tilde A}(\gamma_i,\ \gamma_j)=\{\gamma_s|\gamma_i\preceq \gamma_s\preceq \gamma_i\}$. The set $\mathcal{\tilde A}(\gamma_i,\ \gamma_j)$ is called an inclusive Alexandrov set of $\gamma_i $ and $\gamma_j$.

\begin{law1}\label{L2}
The set $\{\alpha_i|\alpha_i\in\mathcal{G}\}$ is a causal set.

The set $\{\beta_i|\beta_i\in\mathcal{G}\}$ is a causal set.

The set $\{\gamma_i|\gamma_i\in\mathcal{G}\}$ is a causal set.
\end{law1}

Proof: The monad $\alpha_i$ cannot immediately causally precede itself according axiom (\ref{eq:M1.2}). Consequently axiom (\ref{eq:I1.1}) follows from axioms (\ref{eq:M1.2}) and (\ref{eq:M1.7}). The union of two paths is a path iff the last monad of the first path coincides with the first monad of the second path. Consequently axioms (\ref{eq:I1.2}) and (\ref{eq:I1.3}) follows from this properties of paths and axiom (\ref{eq:M1.7}). Axiom (\ref{eq:I1.4}) follows from axioms (\ref{eq:M1.8}). The proof of the second and third sentences is the same. $\Box$

Consequently the considered model of d-graph is a particular case of a causal set.

\begin{law2}\label{T1}
If two monads are connected by an immediate causal priority theirs Alexandrov set is empty.
\end{law2}

Proof: Two monads are connected by an immediate causal priority if they are included in the same edge or x-structure. The case of edge is trivial. Consider the second case. Denote these monads by $\alpha_i$ and $\beta_j$. They are included in $\mathcal{X}_j=\{\alpha_i, \alpha_k, \beta_j, \beta_s\}$. If $\mathcal{A}(\beta_j,\ \alpha_i)$ is not empty it includes some monad $\gamma_m$. We have $\beta_j\prec\gamma_m$ and $\alpha_i\nprec\gamma_m$. Consequently $\alpha_k\prec\gamma_m$. We have $\gamma_m\prec\alpha_i$ and $\gamma_m\nprec\beta_j$. Consequently $\gamma_m\prec\beta_s$. $\alpha_k\prec\beta_s$ by transitivity. But $\beta_s\prec\alpha_k$ in x-structure. This contradiction proves the theorem. $\Box$

Two monads are related by the immediate causal priority if and only if they are causally connected and theirs Alexandrov set is empty. In this model the immediate causal priority or causal connection can be considered as a primordial property or as a consequence.

\begin{law1}\label{L3}
Consider a d-graph $\mathcal{G}$. Let $\hat T$ be the isomorphism from $\mathcal{G}$ to $\mathcal{G}_T$ such that $\hat T\alpha_i=\beta_i$, $\hat T\beta_j=\alpha_j$ and $\hat T(\gamma_i\ \gamma_j)=(\gamma_j\ \gamma_i)$; then $\mathcal{G}_T$ is a d-graph.
\end{law1}

Proof: The proof is an immediate checking of axioms (\ref{eq:M1.1}) - (\ref{eq:M1.8}). $\Box$

The physical meaning of $\hat T$ is a time inversion.

\begin{law1}\label{L4}
The cardinality $|\{\alpha_i|\alpha_i\in\mathcal{G}\}|=|\{\beta_i|\beta_i\in\mathcal{G}\}|$ is an even number.
\end{law1}

Proof: Each monad is included in a unique x-structure. We can consider $\mathcal{G}$ as a set of x-structures. By $N$ denote the number of x-structures in $\mathcal{G}$. We have $|\{\alpha_i|\alpha_i\in\mathcal{G}\}|=|\{\beta_i|\beta_i\in\mathcal{G}\}|=2N$. $\Box$

The past of the monad is the set of monads, which causally precede this monad. The past of $\gamma_i$ is denoted by $\mathcal{P}(\gamma_i)=\{\gamma_j|\gamma_j\prec\gamma_i\}$. The future of the monad is the set of monads, which causally follow this monad. The future of $\gamma_i$ is denoted by $\mathcal{F}(\gamma_i)=\{\gamma_j|\gamma_j\succ\gamma_i\}$. A monad is called maximal iff its future is an empty set. A monad is called minimal iff its past is an empty set.

\begin{law1}\label{L5}
Any maximal monad is a monad of a type $\alpha$. Any minimal monad is a monad of a type $\beta$. The monad a maximal or a minimal monad iff it is external monad. The number of maximal monads in $\mathcal{G}$ is equal to the number of minimal monads.
\end{law1}

Proof: Each monad is included in a unique x-structure. In the x-structure, there are two monads of a type $\beta$ that are included in the past of both monads of a type $\alpha$. The monad of a type $\alpha$ cannot be a minimal monad. The monad of a type $\alpha$ is a maximal monad iff it is not included in a chronon. In the x-structure, there are two monads of a type $\alpha$ that are included in the future of both monads of a type $\beta$. The monad of a type $\beta$ cannot be a maximal monad. The monad of a type $\beta$ is a minimal monad iff it is not included in a chronon. Each chronon includes one monad of a type $\alpha$ and one monad of a type $\beta$. The number of internal monads of a type $\alpha$ is equal to the number of internal monads of a type $\beta$. By Lemma \ref{L4}, $|\{\alpha_i|\alpha_i\in\mathcal{G}\}|=|\{\beta_i|\beta_i\in\mathcal{G}\}|$, so that the number of external monads of a type $\alpha$ is equal to the number of external monads of a type $\beta$. $\Box$

By definition, two chronons $(\alpha_i\ \beta_j)$ and $(\alpha_r\ \beta_k)$ are causally connected and $(\alpha_i\ \beta_j)\prec(\alpha_r\ \beta_k)$ iff $\beta_j\prec\alpha_r$.

\begin{law1}\label{L6}
The set of chronons $\{(\alpha_i\ \beta_j)| (\alpha_i\ \beta_j)\in\mathcal{G}\}$ is a causal set.
\end{law1}

Proof: The proof is the trivial consequence of the partial order of monads. $\Box$

Consider two x-structures $\mathcal{X}_i=\{\alpha_{i1}, \alpha_{i2}, \beta_{i1}, \beta_{i2}\}$ and $\mathcal{X}_j=\{\alpha_{j1}, \alpha_{j2}, \beta_{j1}, \beta_{j2}\}$. By definition, these two x-structures are causally connected and $\mathcal{X}_i\prec\mathcal{X}_j$ iff $(\alpha_{i1}\prec\beta_{j1})\vee(\alpha_{i1}\prec\beta_{j2})\vee(\alpha_{i2}\prec\beta_{j1})\vee(\alpha_{i2}\prec\beta_{j2})$.

\begin{law1}\label{L7}
The set of x-structures $\{\mathcal{X}_i| \mathcal{X}_i \in\mathcal{G}\}$ is a causal set.
\end{law1}

Proof: The proof is the trivial consequence of the partial order of monads. $\Box$

\begin{law1}\label{L8}
There is the x-structure in the set $\{\mathcal{X}_i| \mathcal{X}_i \in\mathcal{G}\}$ such that this x-structure contains two maximal monads. There is the x-structure in the set $\{\mathcal{X}_i| \mathcal{X}_i \in\mathcal{G}\}$ such that this x-structure contains two minimal monads.
\end{law1}

Proof: The set $\{\mathcal{X}_i| \mathcal{X}_i \in\mathcal{G}\}$ is a finite partially ordered set. This set contains maximal and minimal elements. The maximal x-structure contains two maximal monads. The minimal x-structure contains two minimal monads. $\Box$

Define the isomorphism that takes each x-structure to the vertex of some graph and each chronon to the edge of this graph. We get the directed acyclic graph. All properties of this graph are the properties of the d-graph. We can use both mathematical languages: the d-graph and the directed acyclic graph. But the d-graph is more useful for dynamics \cite{1004.5077}. For this reason the considered set of monads in this model is called a dynamical graph or d-graph.

\begin{figure}
	\centering	
		\includegraphics[trim=8cm 17cm 8cm 6cm]{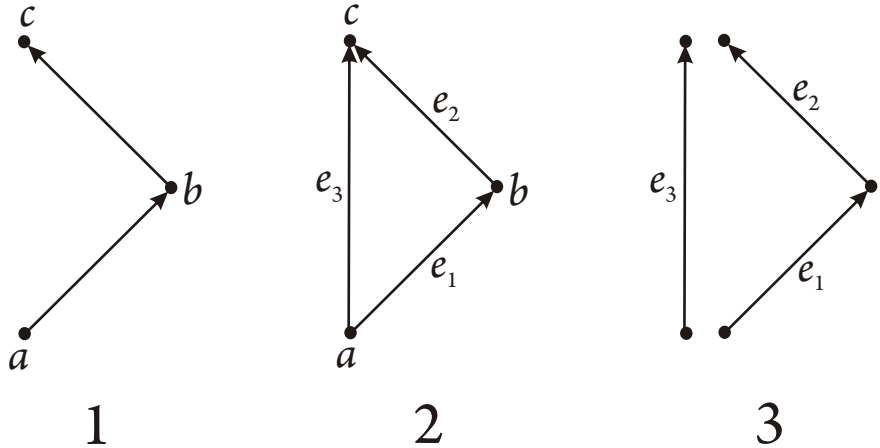}
	\caption{Three directed acyclic graphs.}
	\label{fig:fig5}
\end{figure}

In the case of a d-graph, causal order of monads contain all information about the structure of this d-graph. This is not in the case of a directed acyclic graph. Consider three different graphs (Fig.\ \ref{fig:fig5}).
The causal order of the vertexes is the same for the graphs number 1 and 2: $a\prec b\prec c$. The causal order of the edges is the same for the graphs number 2 and 3: $e_1\prec e_2$, $e_1||e_3$, and $e_2||e_3$.

\section{A SEQUENTIAL GROWTH \label{SG}}

The dynamics of this model is introduced in \cite{1004.5077}. We start from some given d-graph $\mathcal{G}$ and add new x-structures to $\mathcal{G}$ one by one. This procedure is proposed in \cite{Krugly1998,Krugly2002}. Similar procedure and the term `a classical sequential growth dynamics' is proposed in \cite{RideoutSorkin} for another model of causal set.

Any d-graph consists of x-structures. Obviously, we can construct any d-graph by the sequence of the addition of x-structures one by one to the empty set in arbitrary order. In each step, we get the d-graph. It can be a disconnected d-graph. By definition, the d-graph $\mathcal{G}$ is called a connected d-graph iff for any monads $\gamma_i,\ \gamma_j\in\mathcal{G}$ there exists a sequence of immediately causally connected monads with the terminal monads $\gamma_i$ and $\gamma_j$. The simple examples are given in (Fig.\ \ref{fig:fig3}) and (Fig.\ \ref{fig:fig4}). The particular case of the classical sequential growth dynamics \cite{1004.5077} is based on the following Theorem.

\begin{law2}\label{T2}
Consider any connected d-graph $\mathcal{G}_N$ that consists of $N$ x-structures. There exists a sequence of the addition of $N$ x-structures one by one to the empty set such that the following conditions are satisfied:
\begin{itemize}
\item the result of this sequence is $\mathcal{G}_N$;
\item in each step number $N(i)\le N$, the resulting d-graph $\mathcal{G}_{N(i)}$ is a connected d-graph;
\item in each step number $N(i)\le N$, the last added x-structure is a maximal or minimal x-structure in $\mathcal{G}_{N(i)}$.
\end{itemize}
\end{law2}

Proof: Consider the sequence of destructions of $\mathcal{G}_N$. In each step we delete one x-structure. Consider any step number $N(i)\le N$. We must find the maximal or minimal x-structure in $\mathcal{G}_{N(i)}$ such that we get the connected d-graph $\mathcal{G}_{N(i)-1}$ by deleting this x-structure.

Consider an iterative procedure. Take a maximal x-structure $\mathcal{X}_0\in\mathcal{G}_{N(i)}$ and delete it. If the resulting d-graph $\mathcal{G}_{N(i)-1}$ is a connected d-graph consider the next step $N(i)+1$. Otherwise we get two disconnected d-subgraphs $\mathcal{G}_{m(1)}$ and $\mathcal{G}_{k(1)}$. Take a connected d-subgraph $\mathcal{G}_{m(1)}\cup\mathcal{X}_0$. It consists of $m(1)+1$ x-structures. $\mathcal{G}_{m(1)}\cup\mathcal{X}_0$ includes maximal and minimal x-structures. $\mathcal{X}_0$ is a maximal x-structure. Take a minimal x-structure $\mathcal{X}_1\in(\mathcal{G}_{m(1)}\cup\mathcal{X}_0)$ and delete it. If the resulting d-subgraph $(\mathcal{G}_{m(1)}\cup\mathcal{X}_0)\diagdown\mathcal{X}_1$ is a connected d-graph; then $\mathcal{G}_{N(1)}\diagdown\mathcal{X}_1$ is a connected d-graph. Otherwise we divide $\mathcal{G}_{m(1)}\cup\mathcal{X}_0$ into two disconnected d-subgraphs $\mathcal{G}_{m(2)}$ and $\mathcal{G}_{k(2)}$. Assume $\mathcal{X}_0\in\mathcal{G}_{k(2)}$. Take a connected d-subgraph $\mathcal{G}_{m(2)}\cup\mathcal{X}_1$. It includes maximal and minimal x-structures. $\mathcal{X}_1$ is a minimal x-structure. Take a maximal x-structure $\mathcal{X}_2\in(\mathcal{G}_{m(2)}\cup\mathcal{X}_1)$ and delete it. The result is the sequence of d-subgraphs $\mathcal{G}_{m(l)}\subset\dots\subset\mathcal{G}_{m(2)} \subset \mathcal{G}_{m(1)} \subset \mathcal{G}_{N(1)}$. We have $m(l)<\dots < m(2)< m(1)<N(i)$. Either we get the required x-structure in some step or in the last step, we get the d-subgraph $\mathcal{G}_{m(l)}$ that includes only one x-structure $\mathcal{X}_l$. This x-structure is connected only with one x-structure $\mathcal{X}_{l-1}$. Consequently $\mathcal{X}_l$ is a maximal or minimal x-structure in $\mathcal{G}_{N(i)}$, and if we delete $\mathcal{X}_l$ we get the connected d-graph $\mathcal{G}_{N(i)-1}$.

We have the sequence of destructions of $\mathcal{G}_N$. In each step we delete one maximal or minimal x-structure such that the residual d-graph is a connected d-graph by construction. The required sequence of the addition of x-structures is the reverse of the constructed sequence of destructions. $\Box$

\section{ANTICHAINS\label{A}}

A chain is a totally (or a linearly) ordered subset of monads. Every two monads of this subset are related by $\prec$. A chain is a subset of a path. An antichain is a totally unordered subset of monads. Every two elements of this subset are not related by $\prec$. The cardinality of an antichain is called a width of an antichain. A slice is a maximal antichain. Every monad in $\mathcal{G}$ is either in the slice or causal connected to one of its monads. The set of all maximal (or minimal) monads is a slice.

A slice is an important subset of d-graph. The physical meaning is a discrete spacelike hypersurface. Consider the properties of a slice.

\begin{law1}\label{L9}
Any slice $\mathcal{S}\in\mathcal{G}$ divides all monads of $\mathcal{G}$ in three non-overlapping subsets. The first subset is the slice $\mathcal{S}$. The second subsets is a set $\mathcal{B}_1$ such that for any $\gamma_i\in\mathcal{B}_1$ there exists $\gamma_j\in\mathcal{S}$ that $\gamma_i\prec\gamma_j$. The third subsets is a set $\mathcal{B}_2$ such that for any $\gamma_s\in\mathcal{B}_2$ there exists $\gamma_r\in\mathcal{S}$ that $\gamma_r\prec\gamma_s$.
\end{law1}

Proof: Any $\gamma_i\notin\mathcal{S}$ is causal connected with some $\gamma_j\in\mathcal{S}$ by definition of a slice. Suppose there exists $\gamma_0\notin\mathcal{S}$ such that there exist two monads $\gamma_1\in\mathcal{S}$ and $\gamma_2\in\mathcal{S}$ that $\gamma_1\prec\gamma_0$ and $\gamma_0\prec\gamma_2$. We get $\gamma_1\prec\gamma_2$. This contradiction proves the lemma. $\Box$

The subset $\mathcal{B}_1$ is called a past of the slice $\mathcal{S}$ and is denoted by $\mathcal{P}(\mathcal{S})$. The subset $\mathcal{B}_2$ is called a future of the slice $\mathcal{S}$ and is denoted by $\mathcal{F}(\mathcal{S})$.

This property is common for all causal sets. Consider the following specific property of d-graphs. If in a sequence of immediately causally connected monads, each previous monad is a cause of the subsequent monad this sequence is called a directed path or a path. If in a sequence of immediately causally connected monads, each subsequent monad is a cause of the previous monad this sequence is called an opposite directed path. If in a sequence of immediately causally connected monads, some previous monad is a cause of the subsequent monad and some subsequent monad is a cause of the previous monad this sequence is called an undirected path.

\begin{law2}\label{T3} Consider the slice $\mathcal{S}\in\mathcal{G}$, the past $\mathcal{P}(\mathcal{S})$, the future $\mathcal{F}(\mathcal{S})$ of this slice, any monad $\gamma_i\in\mathcal{P}(\mathcal{S})$ and, any monad $\gamma_j\in\mathcal{F}(\mathcal{S})$. If undirected path $\mathcal{UP}$ includes $\gamma_i$ and $\gamma_j$, then there exists a monad $\gamma_0\in\mathcal{UP}$ that $\gamma_0\in\mathcal{S}$. If directed path $\mathcal{DP}$ includes $\gamma_i$ and $\gamma_j$, then there exists an unique monad $\gamma_0\in\mathcal{DP}$ that $\gamma_0\in\mathcal{S}$.
\end{law2}

Proof: Assume the converse. Then any monad of $\mathcal{UP}$ is included either in $\mathcal{P}(\mathcal{S})$ or in $\mathcal{F}(\mathcal{S})$. Consider two successive monads $\gamma_1$ and $\gamma_2$ such that $\gamma_1\in\mathcal{P}(\mathcal{S})$ and $\gamma_2\in\mathcal{F}(\mathcal{S})$. We have $(\gamma_1\ \gamma_2)$ or $(\gamma_2\ \gamma_1)$. There exist $\gamma_3\in\mathcal{S}$ and $\gamma_4\in\mathcal{S}$ such that $\gamma_1\prec\gamma_3$ and $\gamma_4\prec\gamma_2$. If $(\gamma_2\ \gamma_1)$, then $\gamma_4\prec\gamma_3$. This is the contradiction. We have $(\gamma_1\ \gamma_2)$. If this is an edge, then $\gamma_4\prec\gamma_1$ and $\gamma_2\prec\gamma_3$. We get $\gamma_4\prec\gamma_3$. This is the contradiction. If $(\gamma_1\ \gamma_2)$ is included in the x-structure this x-structure includes two monads $\gamma_5$ and $\gamma_6$. We have $(\gamma_5\ \gamma_6)$, $(\gamma_1\ \gamma_6)$, and $(\gamma_5\ \gamma_2)$, then $\gamma_2\prec\gamma_3$ or $\gamma_6\prec\gamma_3$, and $\gamma_4\prec\gamma_1$ or $\gamma_4\prec\gamma_5$. In any case we get $\gamma_4\prec\gamma_3$. Consequently there exists a monad $\gamma_0\in\mathcal{UP}$ that $\gamma_0\in\mathcal{S}$. In the path $\mathcal{DP}$, all monads are causally connected, then the path can include only one monad of an antichain. $\Box$

This property possesses a clear physical meaning. Any slice includes a unique spacelike section of each physical process that starts in the past, and ends in the future of this slice. This property is valid for the slice of edges and monads (halves of edges). This is not true for a slice of x-structures (or vertexes \cite{gr-qc/0506133}).

Two antichains $\mathcal{B}_1$ and $\mathcal{B}_2$ are called ordered antichains, and are denoted $\mathcal{B}_1\prec\mathcal{B}_2$ if there exists the pair of monads $\gamma_1\in\mathcal{B}_1$ and $\gamma_2\in\mathcal{B}_2$ such that $\gamma_1\prec\gamma_2$ and does not exist the pair of monads $\gamma_3\in\mathcal{B}_1$ and $\gamma_4\in\mathcal{B}_2$ such that $\gamma_4\prec\gamma_3$.

The next Theorem is truth for more general case of a causal set than a d-graph.

\begin{law2}\label{T4}
Consider a causal set $\mathcal{C}$. Suppose the cardinality of any slice is finite; then for any slice $\mathcal{S}_0$ there exists a set of slices $\{\mathcal{S}_i\}$ such that this set satisfies the following properties:
\begin{itemize}
\item $\mathcal{S}_0\in\{\mathcal{S}_i\}$;
\item $\{\mathcal{S}_i\}$ is a linearly ordered set;
\item $\cup\{\mathcal{S}_i\}$=$\mathcal{C}$.
\end{itemize}
\end{law2}

Proof: Construct the linearly ordered sequence of slices $\{\mathcal{S}_{i(f)}\}$ such that $\cup\{\mathcal{S}_{i(f)}\}= \mathcal{F}(\mathcal{S}_0)$. Suppose $\mathcal{F}(\mathcal{S}_0)\ne\emptyset$. Otherwise $\mathcal{S}_0$ is a maximal element of $\{\mathcal{S}_i\}$. Take an element $x_1\in\mathcal{F}(\mathcal{S}_0)$ such that $\mathcal{P}(x_1)\cap\mathcal{F}(\mathcal{S}_0)=\emptyset$. We can do this by a local finiteness of $\mathcal{C}$. If $\mathcal{P}(x_1)\cap\mathcal{F}(\mathcal{S}_0)\ne\emptyset$ we take any element of this set. We get the needed element by the finite number of iterative repetitions. Consider $\mathcal{S}_{01}=\mathcal{P}(x_1)\cap\mathcal{S}_0$ and $\mathcal{S}_{02}=\mathcal{S}_0 \setminus\mathcal{S}_{01}$. The set $x_1\cup\mathcal{S}_{02}$ is an antichain. By $\mathcal{M}_0$ denote the set of minimal elements of the set $\mathcal{F}(\mathcal{S}_0) \setminus (\mathcal{F}(x_1)\cup\mathcal{F}(\mathcal{S}_{02}))$. The set $\mathcal{S}_1=x_1\cup\mathcal{S}_{02}\cup\mathcal{M}_0$ is obviously a slice. We have $\mathcal{F}(\mathcal{S}_0)\cap\mathcal{P}(\mathcal{S}_1) =\emptyset$. Take an element $x_i\in\mathcal{F}(\mathcal{S}_{02})$. Similarly, construct a slice $\mathcal{S}_2$ if $\mathcal{P}(x_i)\cap\mathcal{F}(\mathcal{S}_1)=\emptyset$. If $\mathcal{P}(x_i)\cap\mathcal{F}(\mathcal{S}_1)\ne\emptyset$ consider any element of this set. We get the needed element $x_2$ by the finite number of iterative repetitions. We have $\mathcal{P}(x_2)\cap\mathcal{F}(\mathcal{S}_1)=\emptyset$. Construct a slice $\mathcal{S}_2\ni x_2$. Consider the element $x_i$ again. We have $|\mathcal{P}(x_i)\cap\mathcal{F}(\mathcal{S}_1)|>|\mathcal{P}(x_i)\cap\mathcal{F}(\mathcal{S}_2)|$ by a local finiteness of $\mathcal{C}$. We construct a slice $\mathcal{S}_{i-1}$ by the finite number of iterative repetitions such that $\mathcal{P}(x_i)\cap\mathcal{F}(\mathcal{S}_{i-1})=\emptyset$. Construct a slice $\mathcal{S}_i\ni x_i$. Consider $\mathcal{S}_{0i}=\mathcal{S}_0\cap\mathcal{S}_i$. We have $|\mathcal{S}_{0i}|<|\mathcal{S}_{02}|$ by construction. Consider an element $x_j\in\mathcal{F}(\mathcal{S}_{0i})$. The cardinality of any slice is finite, then we can construct a slice $\mathcal{S}_k\in\mathcal{F}(\mathcal{S}_0)$ by the finite number of steps. If the set $\mathcal{F}(\mathcal{S}_0)$ is infinite the construction of the linearly ordered set of slices is an infinite procedure.

Let prove that any element $x_s\in\mathcal{F}(\mathcal{S}_0)$ belongs to some slice with finite number. The element $x_s$ cannot be between two slices $\mathcal{S}_{s-1}$ and $\mathcal{S}_s$ because $\mathcal{P}(\mathcal{S}_s)\cap\mathcal{F}(\mathcal{S}_{s-1})=\emptyset$ by construction. Let prove that $x_s$ cannot be in the future of infinite number of slices of $\{\mathcal{S}_{i(f)}\}$. Consider $\mathcal{P}(x_s)\cap\mathcal{F}(\mathcal{S}_0)$. This is a finite set by a local finiteness of $\mathcal{C}$. We have $|\mathcal{P}(x_s)\cap\mathcal{F}(\mathcal{S}_k)|<|\mathcal{P}(x_s)\cap\mathcal{F}(\mathcal{S}_0)|$ by construction. We get a slice $\mathcal{S}_m$ such that $\mathcal{P}(x_s)\cap\mathcal{F}(\mathcal{S}_m)=\emptyset$. Construct a slice $\mathcal{S}_t\in\mathcal{F}(\mathcal{S}_m)$. We have $x_s\notin\mathcal{F}(\mathcal{S}_t)$ by construction. Either $x_s\in\mathcal{S}_t$ or $x_s\in\mathcal{S}_r$, where $m<r<t$. Similarly, construct the linearly ordered sequence of slices that precede $\mathcal{S}_0$. The union of these two sequences and $\mathcal{S}_0$ is $\{\mathcal{S}_i\}$ $\Box$

The set $\{\mathcal{S}_i\}$ is called a full linearly ordered sequence of slices. We can describe the universe as a linearly ordered sequence of spacelike hypersurfaces by this Theorem. The following Lemma is a consequence of this Theorem.

\begin{law1}\label{L10} Suppose the causal set $\mathcal{C}$ satisfy the conditions of Theorem \ref{T4}; then for any two ordered slices ($\mathcal{S}_1\prec\mathcal{S}_2$) there exists an ordered set of slices $\{\mathcal{S}_i\}$ such that $\mathcal{S}_1\cup\mathcal{S}_2\cup(\mathcal{F}(\mathcal{S}_1)\cap\mathcal{P}(\mathcal{S}_2))=\{\mathcal{S}_i\}$.
\end{law1}

Proof: The set $\mathcal{S}_1\cup\mathcal{S}_2\cup(\mathcal{F}(\mathcal{S}_1)\cap\mathcal{P}(\mathcal{S}_2))$ is a causal set and satisfies the conditions of Theorem \ref{T4}. $\Box$

We can describe any process as an evolution from an initial state to a final state by this Lemma.

\begin{figure}
	\centering	
		\includegraphics[trim=8cm 18cm 8cm 4cm]{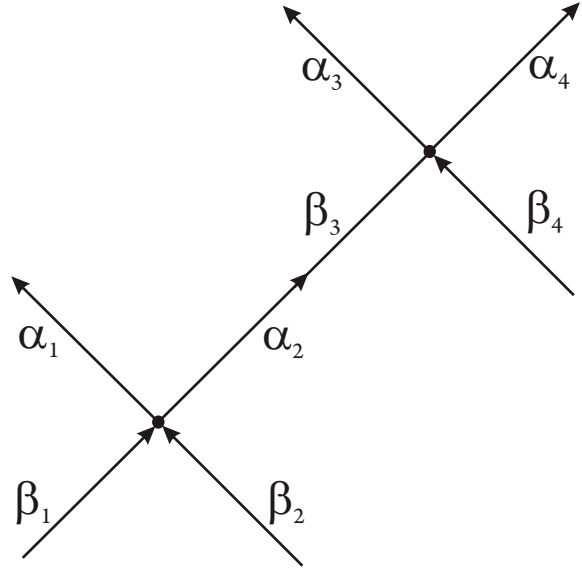}
	\caption{Four ordered slices $\{\beta_1, \beta_2, \beta_4\}$, $\{\alpha_1, \alpha_2, \beta_4\}$, $\{\alpha_1, \beta_3, \beta_4\}$, and $\{\alpha_1, \alpha_3, \alpha_4\}$.}
	\label{fig:fig6}
\end{figure}

In general case, we cannot choose the linearly ordered sequence of slices $\{\mathcal{S}_i\}$ as non-overlapping sets. For example, consider the simple d-graph that consists of two x-structures (Fig.\ \ref{fig:fig6}).
There exists a unique full linearly ordered sequence of slices. These slices are overlapping sets.

\begin{figure}
	\centering	
		\includegraphics[trim=8cm 18cm 8cm 4cm]{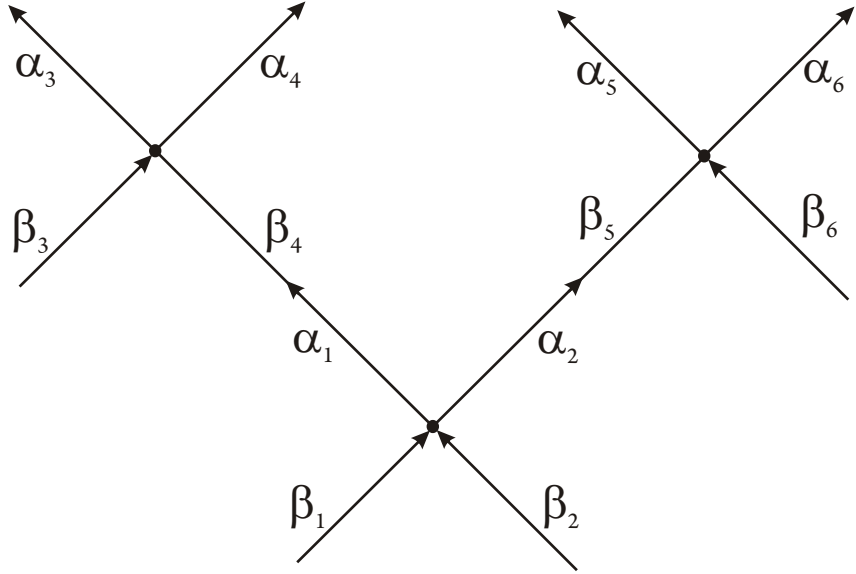}
	\caption{Two non-overlapping slices $\{\beta_3, \beta_4, \alpha_5, \alpha_6\}$ and $\{\alpha_3, \alpha_4, \beta_5, \beta_6\}$.}
	\label{fig:fig7}
\end{figure}

In special relativity theory if two hyperplanes correspond to two relatively moving inertial observers the part of the first hyperplane is in the future of the second hyperplane and the part of the second hyperplane is in the future of the first hyperplane. There are such slices in the d-graph. But in the d-graph such slices can be non-overlapping sets. For example, consider the simple d-graph that consists of three x-structures (Fig.\ \ref{fig:fig7}).

\begin{law1}\label{L11} Suppose $\mathcal{G}$ is a d-graph, $\mathcal{S}$ is a slice, $\mathcal{G}_0$ is a d-subgraph, $\mathcal{M}_1$ is a set of minimal monads of $\mathcal{G}_0$, and $\mathcal{M}_2$ is a set of maximal monads of $\mathcal{G}_0$; then the set $\mathcal{S}_1=\mathcal{S}_0\cup\mathcal{M}_{01}\cup\mathcal{M}_{02}$ is a slice in $\mathcal{G}_0$ where $\mathcal{S}_0=\mathcal{S}\cap\mathcal{G}_0$, $\mathcal{M}_{01}=\mathcal{M}_1\cap\mathcal{F}(\mathcal{S})$, and $\mathcal{M}_{02}=\mathcal{M}_2\cap\mathcal{P}(\mathcal{S})$.
\end{law1}

Proof: Let prove that $\mathcal{S}_1$ is an antichain. Take a minimal monad $\gamma_1\in\mathcal{M}_{01}$. There exists a monad $\gamma_2\in\mathcal{S}$ that $\gamma_2\prec\gamma_1$ by assumption. If there exists a monad $\gamma_3\in\mathcal{S}_1$ that $\gamma_1\prec\gamma_3$ we have $\gamma_2\prec \gamma_1\prec\gamma_3$ in $\mathcal{G}$. This is the contradiction. Take a maximal monad $\gamma_4\in\mathcal{M}_{02}$. There exists a monad $\gamma_5\in\mathcal{S}$ that $\gamma_4\prec\gamma_5$ by assumption. If there exists a monad $\gamma_6\in\mathcal{S}_1$ that $\gamma_6\prec\gamma_4$ we have $\gamma_6\prec \gamma_4\prec\gamma_5$ in $\mathcal{G}$. This is the contradiction. Suppose $\gamma_1\prec\gamma_4$, then $\gamma_2\prec \gamma_1\prec\gamma_4\prec \gamma_5$. This is the contradiction. Consequently $\mathcal{S}_1$ is an antichain.

Let prove that $\mathcal{S}_0$ is a slice. Assume the converse. Then there exists a monad $\gamma_7\in\mathcal{G}_0$ that has not causal connection with the monads of $\mathcal{S}_0$, $\mathcal{M}_{01}$, and $\mathcal{M}_{02}$ in $\mathcal{G}_0$, and does not belong to these sets. We have $\gamma_7\notin\mathcal{S}$ because $\gamma_7\in\mathcal{G}_0$ and $\gamma_7\notin\mathcal{S}_0$. In $\mathcal{G}$, there exists a path $\mathcal{DP}$ between $\gamma_7$ and some monads of $\mathcal{S}$ because $\mathcal{S}$ is a slice. Assume this is the path from $\gamma_7$. The case of the path to $\gamma_7$ is similar. There is not the path from $\gamma_7$ to $\mathcal{S}_0$ in $\mathcal{G}_0$ by assumption. Consequently some monads of $\mathcal{DP}$ belong to $\mathcal{G}_0$ and some monads do not belong. Start from $\gamma_7$ and go to the first monad of $\mathcal{DP}$ that does not belong to $\mathcal{G}_0$. The previous monad $\gamma_8$ of $\mathcal{DP}$ is a maximal monad of $\mathcal{G}_0$. This monad belong to $\mathcal{P}(\mathcal{S})$. Consequently $\gamma_8\in\mathcal{M}_{02}$. We have the causal connection between $\gamma_7$ and $\mathcal{M}_{02}$ in $\mathcal{G}_0$. This contradiction proves the theorem. $\Box$

This Lemma is an algorithm of a construction of slices in d-subgraphs. It is truth for any causal set.

\section{A WIDTH OF D-GRAPH\label{W}}

The cardinality of a slice is called a width of this slice.

\begin{law2}\label{T5}
All slices of a d-graph $\mathcal{G}$ have the same width. This width is called a width of $\mathcal{G}$.\end{law2}

\begin{figure}
	\centering	
		\includegraphics[trim=8cm 18cm 8cm 6cm]{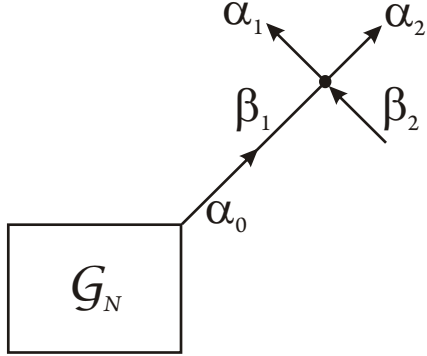}
	\caption{The addition of a new x-structure to one maximal monad.}
	\label{fig:fig8}
\end{figure}

\begin{figure}
	\centering	
		\includegraphics[trim=8cm 18cm 8cm 6cm]{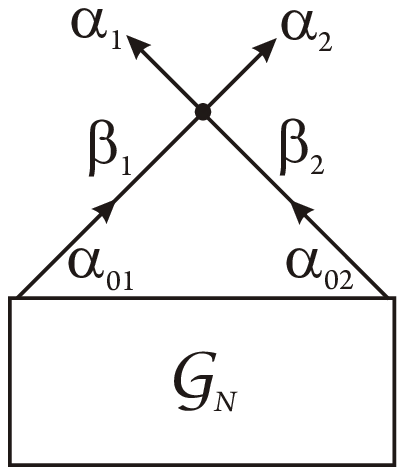}
	\caption{The addition of a new x-structure to two maximal monads.}
	\label{fig:fig9}
\end{figure}

Proof: Denote by $N$ the number of x-structures in $\mathcal{G}_N$. The proof is by induction on $N$. For $N=1$, there is nothing to prove. There are two slices with two monads in each slice. By the inductive assumption, if $\mathcal{G}_N$ consists of $N$ x-structures all slices of $\mathcal{G}_N$ have the same width $n$. We must prove that if $\mathcal{G}_{N+1}$ consists of $N+1$ x-structures all slices of $\mathcal{G}_{N+1}$ have the same width. By Theorem \ref{T2}, we can get any $\mathcal{G}_{N+1}$ by addition of a maximal or minimal x-structure to some $\mathcal{G}_N$. Consider the case of a maximal x-structure. The case of a minimal x-structure is similar. There are two cases: the addition of a new x-structure $\{\beta_1, \beta_2, \alpha_1, \alpha_2\}$ to one maximal monad $\alpha_0$ (Fig.\ \ref{fig:fig8}) and to two maximal monads $\alpha_{01}$ and $\alpha_{02}$ (Fig.\ \ref{fig:fig9}).

Consider the first case. The set $\{\mathcal{S}_i\}$ of all slices in $\mathcal{G}_N$ can be divided in two subsets: $\{\mathcal{S}_{i1}|\alpha_0\in\mathcal{S}_{i1}\}$ and $\{\mathcal{S}_{i2}|\alpha_0\notin\mathcal{S}_{i1}\}$. Consider $\mathcal{S}_{i1}$. We get three slices from $\mathcal{S}_{i1}$ in $\mathcal{G}_{N+1}$. This is $\mathcal{S}_{i1}\cup\beta_2$, $(\mathcal{S}_{i1}\setminus\alpha_0)\cup\beta_1\cup\beta_2$, and $(\mathcal{S}_{i1}\setminus\alpha_0)\cup\alpha_1\cup\alpha_2$. The width of these new slices is $n+1$. Consider $\mathcal{S}_{i2}$. The monad $\alpha_0$ has causal connection with some monads of $\mathcal{S}_{i2}$. Consequently the monads $\beta_1$, $\alpha_1$, and $\alpha_2$ have causal connections with some monads of $\mathcal{S}_{i2}$. We get one slice from $\mathcal{S}_{i2}$ in $\mathcal{G}_{N+1}$. This is $\mathcal{S}_{i2}\cup\beta_2$. The width of this new slice is $n+1$. In $\mathcal{G}_{N+1}$, the width of all slices is $n+1$.

Consider the second case. The set $\{\mathcal{S}_i\}$ of all slices in $\mathcal{G}_N$ can be divided in four subsets: $\{\mathcal{S}_{i1}|\alpha_{01}\in\mathcal{S}_{i1}, \alpha_{02}\in\mathcal{S}_{i1}\}$, $\{\mathcal{S}_{i2}|\alpha_{01}\in\mathcal{S}_{i2}, \alpha_{02}\notin\mathcal{S}_{i2}\}$, $\{\mathcal{S}_{i3}|\alpha_{01}\notin\mathcal{S}_{i3}, \alpha_{02}\in\mathcal{S}_{i3}\}$, and $\{\mathcal{S}_{i4}|\alpha_{01}\notin\mathcal{S}_{i4}, \alpha_{02}\notin\mathcal{S}_{i4}\}$. We get five slices from $\mathcal{S}_{i1}$ in $\mathcal{G}_{N+1}$. This is $\mathcal{S}_{i1}$, $(\mathcal{S}_{i1}\setminus\alpha_{01})\cup\beta_1$, $(\mathcal{S}_{i1}\setminus\alpha_{02})\cup\beta_2$, $(\mathcal{S}_{i1}\setminus(\alpha_{01}\cup\alpha_{01}))\cup\beta_1\cup\beta_2$, and $(\mathcal{S}_{i1}\setminus(\alpha_{01}\cup\alpha_{01}))\cup\alpha_1\cup\alpha_2$. We get two slices from $\mathcal{S}_{i2}$ in $\mathcal{G}_{N+1}$. This is $\mathcal{S}_{i2}$ and $(\mathcal{S}_{i2}\setminus\alpha_{01})\cup\beta_1$. The case of $\mathcal{S}_{i3}$ is the same. We get one slice from $\mathcal{S}_{i4}$ in $\mathcal{G}_{N+1}$. This is $\mathcal{S}_{i4}$. In $\mathcal{G}_{N+1}$, the width of all slices is $n$. $\Box$

This Theorem has a clear physical meaning. This is the conservation law of the number of primordial indivisible objects. If there is the conservation law of the number of primordial objects in each elementary interaction (x-structure) there is the conservation law of the number of primordial objects in each process. This Theorem is the base of the classification of d-graphs and d-subgraphs. We can classify all physical processes in microscopic level by using the number of primordial objects that simultaneously take part in these processes.

\section{AN IDENTIFICATION OF OBJECTS \label{IO}}

In relativity theory, any object is a set of unconnected points at one time instant. Similarly, in discrete mechanics, any object is an antichain. An antichain has one property. This is cardinality. If d-graph describes $k$ objects the slice is the union of $k$ non-intersecting antichains. The interaction of objects is a monad exchange.

\begin{figure}
	\centering	
		\includegraphics[scale=0.8,trim=8cm 4cm 8cm 3cm]{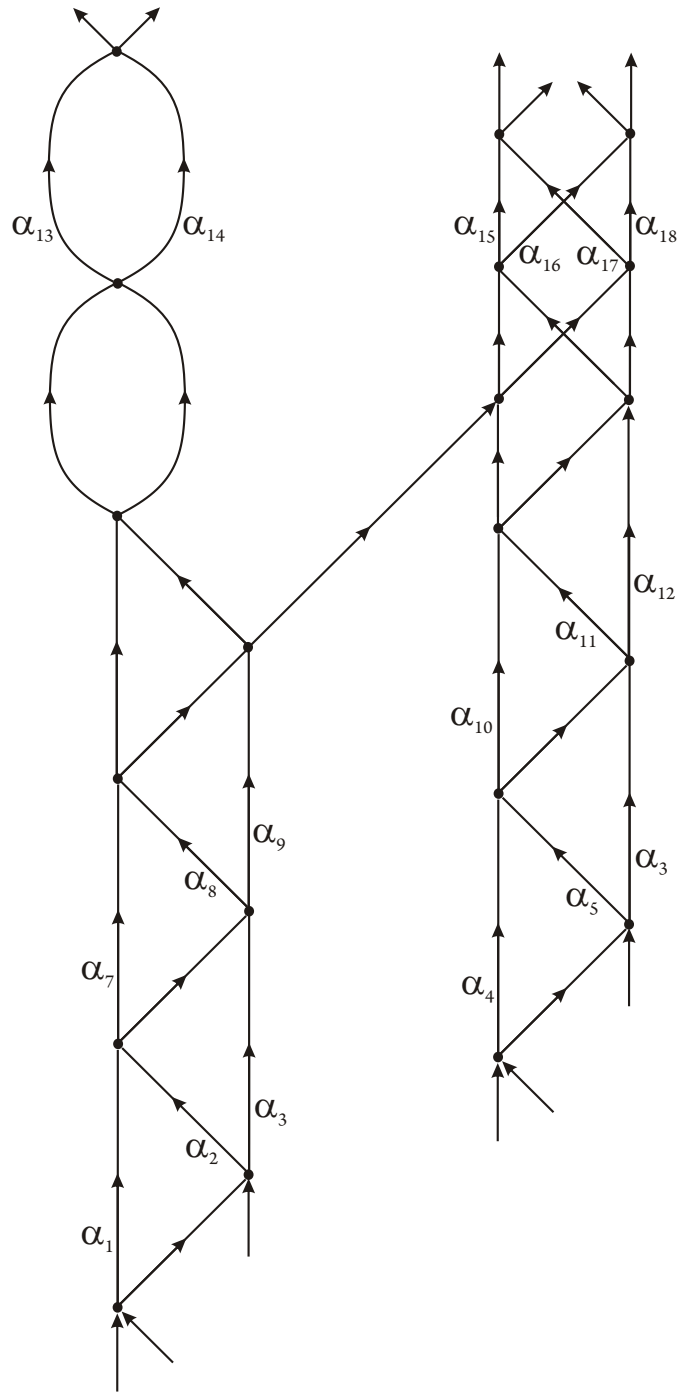}
	\caption{Two interacting objects.}
	\label{fig:fig10}
\end{figure}

Consider a simple example (Fig.\ \ref{fig:fig10}).
Two objects interact. Before interaction each object consists of 3 monads. After interaction the first object consists of 2 monads and the second object consists of 4 monads.

We can describe this interaction using state vectors and creation and destruction operators. Denote these objects before and after interaction by $|3\rangle$, $|3\rangle$, $|2\rangle$, and $|4\rangle$, respectively. We have
\[
|4\rangle =\hat a^{\dag}|3\rangle \textrm{,}
\]
\begin{equation}
\label{eq:IO1.1}
|2\rangle =\hat a|3\rangle \textrm{}
\end{equation}
for the interaction. The creation and destruction operators was used for creation and destruction of the edges of the graph in \cite{DP79}. But this model is undirected graph.

In this simple example the objects are clearly visible. But we must have the formal algorithm to divide the slice in antichains. We cannot get such algorithm if we consider only the slice. We must consider the past and the future of the slice. The simplest criterion is based on a d-graph distance. This is the number of monads in the shortest undirected path between two monads. We can define the antichain as an object if the d-graph distance between any two monads of this antichain is less than the d-graph distance between any monad of this antichain and any other monad of the slice. In our example, such objects are the antichains $\{\alpha_1, \alpha_2, \alpha_3\}$ and $\{\alpha_4, \alpha_5, \alpha_6\}$. In general case, an object can include some monads such that the d-graph distance between these monads and some monads of this object is not less than the d-graph distance between these monads and some other monads of the slice. Such monads can be included in several objects, and these objects have fuzzy boundaries. A slice can form a hierarchy of objects. The d-graph distance between any two monads of any object of the $k$-level is less than the d-graph distance between any monads that are included in the same object of the $(k+1)$-level but are not included in the same object of the $k$-level.

Consider the evolution of objects. There are two ordered slices $\mathcal{S}_1\prec\mathcal{S}_2$. These slices are divided in the objects. The question is: what objects of $\mathcal{S}_2$ are future states of the objects of $\mathcal{S}_1$? The weakest criterion is causal connections of monads. Two objects are two stage of the same object if some monads of the first object are causally connected or coincide with some monads of the second object. In our example, the antichain $\{\alpha_7, \alpha_8, \alpha_9\}$ is the future stage of the antichain $\{\alpha_1, \alpha_2, \alpha_3\}$, and the antichain $\{\alpha_{10}, \alpha_{11}, \alpha_{12}\}$ is not. But in this case, both antichains $\{\alpha_{13}, \alpha_{14}\}$ and $\{\alpha_{15}, \alpha_{16}, \alpha_{17}, \alpha_{18}\}$ are the future stages of the antichain $\{\alpha_1, \alpha_2, \alpha_3\}$. We can use the average d-graph distance to distinguish these antichains. But in this example we have the same distance. We can use some intensity of causal connections. For example, this can be the average number of paths between monads of the previous stage and monads of the future stage. Two objects are two stages of the same object if their monads have the greatest average intensity of causal connections.

\section{A LIGHT CONE \label{LC}}

The physical meaning of the sets $\mathcal{P}(\gamma_i)=\{\gamma_j|\gamma_j\prec\gamma_i\}$ and $\mathcal{F}(\gamma_i)=\{\gamma_j|\gamma_j\succ\gamma_i\}$ is a past light cone and a future light cone of $\gamma_i$, respectively. The cardinality of the antichain $\mathcal{P}(\gamma_i)\cap\mathcal{S}$ (or $\mathcal{F}(\gamma_i)\cap\mathcal{S}$) is called the width of the past (or future) light cone in $\mathcal{S}$.

\begin{law1}\label{L13} Consider the past light cone $\mathcal{P}(\gamma_0)$, the future light cone $\mathcal{F}(\gamma_0)$ and two slices $\mathcal{S}_1$ and $\mathcal{S}_2$ in the d-graph $\mathcal{G}$. If $\mathcal{B}_{p1}\prec\mathcal{B}_{p2}$, where the antichain $\mathcal{B}_{p1}=\mathcal{S}_1\cap\mathcal{P}(\gamma_0)$ and the antichain $\mathcal{B}_{p2}=\mathcal{S}_2\cap\mathcal{P}(\gamma_0)$; then the width of $\mathcal{P}(\gamma_0)$ in $\mathcal{S}_1$ is not less than the width of $\mathcal{P}(\gamma_0)$ in $\mathcal{S}_2$. If $\mathcal{B}_{f1}\prec\mathcal{B}_{f2}$, where the antichain $\mathcal{B}_{f1}=\mathcal{S}_1\cap\mathcal{F}(\gamma_0)$ and the antichain $\mathcal{B}_{f2}=\mathcal{S}_2\cap\mathcal{F}(\gamma_0)$; then the width of $\mathcal{F}(\gamma_0)$ in $\mathcal{S}_1$ is not greater than the width of $\mathcal{F}(\gamma_0)$ in $\mathcal{S}_2$.
\end{law1}

Proof: Let prove this lemma for a past light cone. In this case, the proof for a future light cone is a consequence of Lemma \ref{L3}. Consider a d-subgraph $\mathcal{G}_0$ of $\mathcal{G}$. It consist of all x-structures $\mathcal{X}_i$ such that some monads of $\mathcal{X}_i$ belong to $\mathcal{P}(\gamma_0)$. Obviously, $\mathcal{G}_0=\mathcal{P}(\gamma_0)\cup\mathcal{M}$, where $\mathcal{M}$ is the set of some maximal monads in $\mathcal{G}_0$. Consider slices $\mathcal{S}_{01}\supset\mathcal{B}_{p1}$ and $\mathcal{S}_{02}\supset\mathcal{B}_{p2}$ in $\mathcal{G}_0$. By Lemma \ref{L11}, $\mathcal{S}_{01}=\mathcal{B}_{p1}\cup\mathcal{M}_{01}$ and $\mathcal{S}_{02}=\mathcal{B}_{p2}\cup\mathcal{M}_{02}$, where $\mathcal{M}_{01}\subset\mathcal{M}$ and $\mathcal{M}_{02}\subset\mathcal{M}$. Some monads of $\mathcal{M}_{02}$ can belong to the future light cone of some monads of $\mathcal{B}_{p1}$. But some monads of $\mathcal{M}_{01}$ cannot belong to the future light cone of some monads of $\mathcal{B}_{p2}$. Otherwise, these monads of $\mathcal{M}_{01}$ belong to the future light cone of some monads of $\mathcal{B}_{p1}$ because $\mathcal{B}_{p1}\prec\mathcal{B}_{p2}$. Consequently $\mathcal{M}_{01}\le\mathcal{M}_{02}$. By Theorem \ref{T5}, all slices has the same width. Consequently $\mathcal{B}_{p1}\ge\mathcal{B}_{p2}$. $\Box$

In this model, the horizon of any observer cannot shrinks.

\section{CONCLUSION \label{CON}}

This paper and the paper \cite{1004.5077} introduce a consistent description of the model: a kinematics and a dynamics. However this is an abstract model without any connections with physical phenomena. The goal of this model is to get the theory of particles. In this approach, particles must be identified with some structures of a d-graph in a like manner as in \cite{ STC4, BTSO1} Also, some properties of structures must be identified with physical quantities such as a mass, an energy, a momentum, a charge etc. The properties of particles are considered now as manifestations of symmetry. Consequently the first open question is the investigation of symmetry of structures in a d-graph.

\section*{ACKNOWLEDGEMENTS}
I am grateful to several colleagues for extensive discussions on this subject, especially Alexandr V. Kaganov and Vladimir V. Kassandrov.

\end{document}